\documentclass[twocolumn]{26onr_art}
\usepackage{ulem}
\usepackage{harvard}
\usepackage{graphicx}

\pdfoutput = 1

\def\be{\begin{eqnarray}}
\def\ee{\end{eqnarray}}
\def\benl{\begin{eqnarray*}}
\def\eenl{\end{eqnarray*}}

\newcommand{\nwc}{\newcommand}
\nwc{\bm}{\boldmath}
\nwc{\m}{\mbox}
\nwc{\ubm}{\unboldmath}
\nwc{\bmU}{\m{\bm$U$\ubm}}
\nwc{\bmX}{\m{\bm$X$\ubm}}
\nwc{\bmu}{\m{\bm$u$\ubm}}
\nwc{\bmx}{\m{\bm$x$\ubm}}
\nwc{\bmz}{\m{\bm$z$\ubm}}
\nwc{\bmv}{\m{\bm$v$\ubm}}
\nwc{\bmw}{\m{\bm$w$\ubm}}
\nwc{\bmW}{\m{\bm$W$\ubm}}
\nwc{\bmn}{\m{\bm$n$\ubm}}
\nwc{\bmG}{\m{\bm$G$\ubm}}
\nwc{\bmF}{\m{\bm$F$\ubm}}
\nwc{\bmI}{\m{\bm$I$\ubm}}
\nwc{\bmN}{\m{\bm$N$\ubm}}
\nwc{\bmP}{\m{\bm$P$\ubm}}
\nwc{\bmcalP}{\m{\bm $\cal P$\ubm}}
\nwc{\bmV}{\m{\bm$V$\ubm}}
\nwc{\bmS}{\m{\bm$S$\ubm}}

\begin{document}

\title{The numerical simulation of ship waves using cartesian-grid and volume-of-fluid methods}

\author{Douglas G.\ Dommermuth$^1$, Thomas T. O'Shea$^1$, Donald C. Wyatt$^1$, \\ Mark Sussman$^2$, Gabriel D. Weymouth$^3$, Dick K.P. Yue$^3$, \\ Paul Adams$^4$, and Randall Hand$^4$ }

\affiliation{\small $^1$Naval Hydrodynamics Division, Science Applications International Corporation,
\\ 10260 Campus Point Drive, MS 35, San Diego, CA  92121 \\ $^2$Department of Mathematics,
Florida State University, Tallahassee, FL 32306 \\ $^3$Department of Mechanical Engineering, Massachusetts Institute of Technology, Cambridge, MA 02139\\$^4$U.S. Army Engineer Research and Development Center, Vicksburg, MS 39180}

\maketitle

\begin{abstract}
Cartesian-grid methods in combination with immersed-body and volume-of-fluid methods are ideally suited for simulating breaking waves around ships.  A surface panelization of the ship hull is used as input to impose body-boundary conditions on a three-dimensional cartesian grid.  The volume-of-fluid portion of the numerical algorithm is used to capture the free-surface interface, including the breaking of waves.  The numerical scheme is implemented on a parallel computer.  Various numerical issues are discussed, including implementing exit boundary conditions, conserving mass using a novel regridding algorithm, improving resolution through the use of stretched grids, minimizing initial transients, and enforcing hull boundary conditions on cartesian grids.  Numerical predictions are compared to experimental measurements of ship models moving with forward speed, including model 5415 and model 5365 (Athena).  The ability to model forced-motions is illustrated using a heaving sphere moving with forward speed.
\end{abstract}

\section{Introduction}

The Numerical Flow Analysis (NFA) code provides turnkey capabilities to model breaking waves around a ship, including both plunging and spilling breaking waves, the formation of spray, and the entrainment of air.   NFA uses a cartesian-grid formulation with immersed-body and volume-of-fluid (VOF) methods.  The governing equations are formulated on a cartesian grid thereby eliminating complications associated with body-fitted grids. The sole geometric input into NFA is a surface panelization of the ship hull.  No additional gridding beyond what is already used in potential-flow methods and hydrostatics calculations is required.   The ease of input in combination with a flow solver that is implemented using parallel-computing methods permit the rapid turn around of numerical simulations of complex interactions between free surfaces and ships.   

Based on \citeasnoun{colella99}, free-slip boundary conditions are imposed on the surface of the ship hull.  The fractional areas and volumes of very small cells are merged to improve the conditioning of the Poisson solver \cite{mampaey95}.  The grid is stretched along the cartesian axes using one-dimensional elliptic equations to improve resolution near the ship hull and the free-surface interface.    Away from the ship and the free-surface interface, where the flow is less complicated, the mesh is coarser. Details of the grid stretching algorithm, which uses weight functions that are specified in physical space, are provided in \citeasnoun{knupp93}.   Free-slip boundary conditions are also imposed at the entrance, along the sides, and at the top and bottom of the computational domain.   At the exit, Orlanski-like boundary conditions are imposed \cite{orlanski93}.  

The VOF portion of the numerical algorithm is used to track the free-surface interface, including the large-scale effects of breaking waves, spray formation and air entrainment.   A novel regridding scheme is introduced whereby the level of water at the entrance of the computational domain is preserved. The interface tracking of the free surface is second-order accurate in space and time.   At each time step, the position of the free surface is reconstructed using piece-wise planar surfaces \cite{rider94,gueyffier99}. The advection portion of the VOF algorithm uses an operator-split method \cite{puckett97}.  The advection algorithm implements a correction to improve mass conservation when the flow is not solenoidal due to numerical errors.   

The convective terms in the momentum equations are treated using a slope-limited, third-order QUICK scheme as discussed in \citeasnoun{leonard97}.   A Smagorinsky turbulence model is also implemented.. There are no special treatments required to model either the flow separation at the transom or the wave overturning at the bow.   A second-order, variable-coefficient Poisson equation is used to project the velocity onto a solenoidal field. A preconditioned conjugate-gradient method is used to solve the Poisson equation. 

NFA is written in Fortran 90.   The governing equations are solved using a domain-decomposition method.   The domains are distributed over the nodes of a parallel computer.  Communication between processors on the Cray XT3 and X1 is performed using either CrayÕs shared memory access library (SHMEM) or MPI.  The CPU requirements are linearly proportional to the number of grid points and inversely proportional to the number of processors.  For the most part, NFA performed equally well on the Cray X1, which is a vector machine, and the Cray XT3, which is a massively parallel machine.  The main exception involved portions of NFA that involved conditional statements that could not be vectorized on the Cray X1.  Together, the ease of input and usage, the ability to model and resolve complex free-surface phenomena, and the speed of the numerical algorithm provide a robust capability for simulating the free-surface disturbances near a ship.

\section{\label{sec:formulation}Formulation}

Consider turbulent flow at the interface between air and water.  Let $u_i$ denote the three-dimensional velocity field as a function of space ($x_i$) and time ($t$).  The coordinate system is fixed with respect to the ship.   $v_i$ is the velocity of the ship.  $v_i$ includes the effects of rigid-body translation and rigid-body rotation.  For an incompressible flow, the conservation of mass gives
\begin{eqnarray}
\label{mass}
\frac{\partial u_i}{\partial x_i} = 0 \;\; .
\end{eqnarray}
\noindent  $u_i$ and $x_i$ are normalized by $U_o$ and $L_o$, which denote the free-stream velocity and the length of the body,
respectively.

Following a procedure that is similar to \citeasnoun{rider94}, we let $\phi$ denote the fraction of fluid that is inside a cell. By
definition, $\phi=0$ for a cell that is totally filled with air, and $\phi=1$ for a cell that is totally filled with water.

The advection of $\phi$ is expressed as follows:
\begin{eqnarray}
\label{vof}
\frac{\partial \phi}{\partial t}+ \frac{\partial}{\partial x_j} \left[ (u_j-v_j)\phi \right]= \frac{\partial Q}{\partial x_j} \;\; ,
\end{eqnarray}
\noindent  The coordinate system is fixed with respect to the body \cite{repetto}.  $Q$ is a sub-grid-scale flux which can model the entrainment of gas into the liquid.  \citeasnoun{dommermuth98} provide details of a sub-grid model that is appropriate for interface capturing methods that allow mixing of air and water.   Since the present formulation maintains a sharp interface, $Q=0$.

Let $\rho_\ell$ and $\mu_\ell$ respectively denote the density and dynamic viscosity of water. Similarly, $\rho_g$ and $\mu_g$ are the corresponding properties of air.  The flows in the water and the air are governed by the Navier-Stokes equations:
\begin{eqnarray}
\label{navi}
\frac{d u_i}{d t}+\frac{\partial}{\partial x_j} \left[ (u_j-v_j) u_i \right]  =  -\frac{1}{\rho} \frac{\partial P}{\partial x_i}  \nonumber  \\
+\frac{1}{\rho R_e} \frac{\partial}{\partial x_j} \left( 2 \mu S_{ij} \right) -\frac{G_i}{F_r^2}  +\frac{\partial  \tau_{ij}}{\partial x_j} \;\; ,
\end{eqnarray}
\noindent where $R_e=\rho_\ell U_o L_o/\mu_\ell$ is the Reynolds number and $F_r^2 = U_o^2/(g L_o)$ is the Froude number. $g$ is the acceleration of gravity. $G_i$ is a body force that rotates with the body \cite{repetto}. $P$ is the pressure.  As described in \citeasnoun{dommermuth98}, $\tau_{ij}$ is the subgrid-scale stress tensor. $S_{ij}$ is the deformation tensor:
\begin{eqnarray}
S_{ij} & = & \frac{1}{2} \left( \frac{\partial u_i}{\partial x_j} +\frac{\partial u_j}{\partial x_i} \right) \;\; .
\end{eqnarray}
\noindent $\rho$ and $\mu$ are respectively the dimensionless variable densities and viscosities:
\begin{eqnarray}
\label{density}
\rho(\phi) & = & \lambda + (1 - \lambda) {\rm H} (\phi) \nonumber \\ \mu(\phi) &
= & \eta + (1 - \eta ) {\rm H} (\phi) \;\; ,
\end{eqnarray}
\noindent where $\lambda = \rho_g/\rho_\ell$ and $\eta = \mu_g/\mu_\ell$ are the density and viscosity ratios between air and water.
For a sharp interface, with no mixing of air and water, ${\rm H}$ is a step function.  In practice, a mollified step function is used to
provide a smooth transition between air and water.

A no-flux condition is imposed on the surface of the ship hull:
\begin{eqnarray}
\label{eqn:neumann}
u_i n_i = v_i n_i
\end{eqnarray}
\noindent where $n_i$ denotes the normal to the ship hull that points into the fluid.

As discussed in \citeasnoun{dommermuth98}, the divergence of the momentum equations (\ref{navi}) in combination with the
conservation of mass (\ref{mass}) provides a Poisson equation for the dynamic pressure:
\begin{eqnarray}
\label{pois} 
\frac{\partial}{\partial x_i} \frac{1}{\rho} \frac{\partial
P}{\partial x_i} = \Sigma \;\; ,
\end{eqnarray}
\noindent where $\Sigma$ is a source term.  As shown in the next section, the pressure is used to project the velocity onto a
solenoidal field.

\subsection{Numerical Time Integration}

Based on \citeasnoun{sussman03a}, a second-order Runge-Kutta scheme is used to integrate with respect to time the field equations
for the velocity field.  Here, we illustrate how a volume of fluid formulation is used to advance the volume-fraction function.  Similar examples are provided by \citeasnoun{rider94}.  During the first stage of the Runge-Kutta algorithm, a Poisson equation for the pressure
is solved:
\begin{eqnarray}
\label{eqn:poisson1}
\frac{\partial}{\partial x_i} \frac{1}{\rho(\phi^k)} \frac{\partial P^*}{\partial x_i} =\frac{\partial}{\partial x_i} \left(
\frac{u^k_i}{\Delta t}+R_i \right) \;\; ,
\end{eqnarray}
where $R_i$ denotes the nonlinear convective, hydrostatic, viscous, sub-grid-scale, and body-force terms in the momentum equations.
$u^k_i$ and $\rho^k$ are respectively the velocity components at time step $k$.  $\Delta t$ is the time step.  $P^*$ is the
first prediction for the pressure field.

For the next step, this pressure is used to project the velocity onto a solenoidal field. The first prediction for the
velocity field ($u^*_i$) is
\begin{eqnarray}
\label{eqn:runge1}
u^*_i=u^k_i+\Delta t \left( R_i-\frac{1}{\rho(\phi^k)}\frac{\partial P^*}{\partial x_i} \right)
\end{eqnarray}
The volume fraction is advanced using a volume of fluid operator (VOF):
\begin{eqnarray}
\phi^*=\phi^{k}- {\rm VOF} \left( u^k_i,\phi^k,\Delta t \right)
\end{eqnarray}
Details of the VOF operator are provided later.  A Poisson equation for the pressure is solved again during the second stage of the Runge-Kutta algorithm:
\begin{eqnarray}
\label{eqn:poisson2}
\frac{\partial}{\partial x_i} \frac{1}{\rho(\phi^*)} \frac{\partial P^{k+1}}{\partial x_i}=\frac{\partial}{\partial x_i} \left(
\frac{u^*_i+u^k_i}{\Delta t}+R_i \right)
\end{eqnarray}
$u_i$ is advanced to the next step to complete one cycle of the Runge-Kutta algorithm:
\begin{eqnarray}
\label{eqn:runge2}
u^{k+1}_i=\frac{1}{2} \left( u^*_i + u^k_i +\Delta t \left( R_i -\frac{1}{\rho(\phi^*)}\frac{\partial P^{k+1}}{\partial x_i}
\right) \right) \;\; ,
\end{eqnarray}
and the volume fraction is advanced to complete the algorithm:
\begin{eqnarray}
\phi^{k+1}=\phi^k- {\rm VOF} \left( \frac{u^*_i+u^k_i}{2},\phi^{k},\Delta t \right)
\end{eqnarray}

\subsection{Gridding}

Along the cartesian axes, one-dimensional stretching is performed  using a differential equation.    Let $x$ denote the position of the grid points in physical space, and let $\xi$ denote the position of the grid points in a mapped space.   As shown by \citeasnoun{knupp93}, the differential equation that describes grid stretching in one dimension is as follows:
\begin{eqnarray}
\label{eqn:grid}
\frac{\partial^2 x}{\partial \xi^2} + \frac{1}{w} \frac{\partial w}{\partial \xi} \frac{\partial x}{\partial \xi} = 0 \;\; ,
\end{eqnarray}
\noindent where $w(x)$ is a weight function that is specified in physical space.   For example, suppose the grid spacing is constant but different for $x<x_o$ and $x>x_1$.    Between $x_o \leq x \leq x_1$, there is a transition zone from one grid spacing to the next.    Then the following weight function may be used to describe this distribution of grid points:
\begin{eqnarray}
w(x) & = & w_0 \;\; {\rm for} \;\;  x < x_0 \nonumber \\
w(x) & = & \frac{w_0-w_1}{2} \left(1+\cos(\frac{\pi (x-x_0)}{x_1-x_0}) \right) \nonumber \\
         & + & w_1 \;\; {\rm for} \;\; x_0 \leq x \leq  x_1 \nonumber \\
w(x) & = & w_1 \;\; {\rm for} \;\; x > x_1  \;\; .
\end{eqnarray}
\noindent Using this approach, multiple zones of grid clustering may be specified.   For example, along the x-axis ($x_1$ in indical notation), grid points may be clustered near the bow and stern.    For the y-axis ($x_2$ in indical notation), grid points are clustered near the centerline out beyond the half beam.     Finally, for the z-axis ($x_3$ in indical notation), grid points are clustered near the mean waterline in a region that is between the top and bottom of the ship hull.   Note that equation \ref{eqn:grid}, is a nonlinear equation that is solved iteratively.

\subsection{Enforcement of Body Boundary Conditions}

A no-flux boundary condition is imposed on the surface of the body using a finite-volume technique.  A signed distance function $\psi$ is used to represent the body.  $\psi$ is positive outside the  body and negative inside the body. The magnitude of $\psi$ is the minimal distance between the position of $\psi$ and the surface of the body.  $\psi$ is zero on the surface of the body.   $\psi$ is calculated using a surface panelization of the hull form.   Green's theorem is used to indicate whether a point is inside or outside the body, and then the shortest distance from the point to the surface of the body is calculated.  Triangular panels are used to discretize the surface of the body.  The shortest distance to the surface of the body can occur on either a surface, edge, or vertice of a triangular panel.  Details associated with the calculation of $\psi$ are provided in \citeasnoun{sussman01}.

Cells near the ship hull may have an irregular shape, depending on how the surface of the ship hull cuts the cell.  On these irregular boundaries, the finite-volume approach is used to impose free-slip boundary conditions.   Let $S_b$ denote the portion of the cell whose surface is on the body, and let $S_o$ denote the other bounding surfaces of the cell that are not on the body.   Gauss's theorem is applied to the volume integral of  equation \ref{eqn:poisson1}:
\begin{eqnarray}
\label{eqn:integral1}
\int_{S_o+S_b} ds  \frac{n_i}{\rho(\phi^k)} \frac{\partial P^*}{\partial x_i}  = \int _{S_o+S_b} ds \left( \frac{u^k_i n_i}{\Delta t}+R_i n_i \right) \;\; .
\end{eqnarray}
\noindent Here, $n_i$ denotes the components of the unit normal on the surfaces that bound the cell.   Based on equation \ref{eqn:runge1}, a Neumann condition is derived for the pressure on $S_b$ as follows:  
\begin{eqnarray}
\label{eqn:bc1}
\frac{n_i}{\rho(\phi^k)}\frac{\partial P^*}{\partial x_i} =-\frac{u^*_i n_i}{\Delta t} +\frac{u^k_i n_i}{\Delta t}+R_i n_i \;\; .
\end{eqnarray}
\noindent The Neumann condition for the velocity (\ref{eqn:neumann}) is substituted into the preceding equation to complete the Neumann condition for the pressure on $S_b$:
\begin{eqnarray}
\label{eqn:bc2}
\frac{n_i}{\rho(\phi^k)}\frac{\partial P^*}{\partial x_i} =-\frac{v^*_i n_i}{\Delta t} +\frac{u^k_i n_i}{\Delta t}+R_i n_i \;\; .
\end{eqnarray}
\noindent This Neumann condition for the pressure is substituted into the integral formulation in equation \ref{eqn:integral1}:
\begin{eqnarray}
\label{eqn:integral2}
\int_{S_o} ds  \frac{1}{\rho(\phi^k)} \frac{\partial P^*}{\partial x_i} n_i & = & \int _{S_o} ds \left( \frac{u^k_i n_i}{\Delta t}+R_i n_i \right) \nonumber \\
& + & \int _{S_b} ds \frac{v^*_i n_i}{\Delta t}
\end{eqnarray}
This equation is solved using the method of fractional areas.  Details associated with the calculation of the area fractions are provided in \citeasnoun{sussman01} along with additional references.  Cells whose cut volume is less than 25\% of the full volume of the cell are merged with neighbors.  The merging occurs along the direction of the steepest gradient of the signed-distance function $\psi$. This improves the conditioning of the Poisson equation for the pressure.   As a result, the stability of the  projection operator for the velocity is also improved (see equations \ref{eqn:runge1} and \ref{eqn:runge2}). 

\subsection{Interface reconstruction and advection}

In our VOF formulation, the free surface is reconstructed from the volume fractions using piece-wise linear polynomials.  The reconstruction is based on algorithms that are described by \citeasnoun{gueyffier99}. The surface normals are estimated using weighted central differencing of the volume fractions. A similar algorithm is described by \citeasnoun{pilliod97}.     Near the body, care must be taken to use cells whose volume fraction is exterior to the body in the calculation of the normal to the free-surface interface. The advection portion of the algorithm is operator split, and it is based on similar algorithms reported in \citeasnoun{puckett97}.   Major differences between the present algorithm and earlier methods include special treatments to account for the body and to alleviate mass-conservation errors due to the presence of non-solenoidal velocity fields.

Let $F_i$ denote flux through the faces of a cell.   $F_i$ is expressed in terms of the relative velocity ($u_i-v_i$) and the  areas of the faces of the cell  ($A_i$) that are cut by the ship hull:
\begin{eqnarray}
F_i = A_i \left( u_i - v_i \right)  \;\; .
\end{eqnarray}
\noindent If the ship hull does not cut the cell, then $A_i$ correspond to the surface areas that bound the cell.   Near the ship hull, $A_i$ is some fraction of the surface areas that bound the cell.   Note that $A_i=0$ inside the ship hull.  Based on an application of Gauss's theorem to the volume integral of Equation \ref{mass} and making use of Equation \ref{eqn:neumann}:
\begin{eqnarray}
\label{eqn:diverge}
F^+_i-F^-_i = 0 \;\; ,
\end{eqnarray}
\noindent where $F^+_i$ is the flux on the positive i-th face of the cell and $F^-_i$ is the flux on the negative i-th face of the cell.   Due to numerical errors, equation \ref{eqn:diverge} is not necessarily satisfied.   Let $\cal E$ denote the resulting numerical error for any given cell.   For each cell whose flux is not conserved,  a correction is applied prior to performing the VOF advection.   For example,  the following reassignment of the flux along the vertical direction ensures that the redefined flux is conserved:
\begin{eqnarray}
\tilde{F}^+_3 & = & F^+_3 - \frac{{\cal E}A^+_3}{A^+_3+A^-_3} \nonumber \\
\tilde{F}^-_3 & = & F^-_3  + \frac{{\cal E}A^-_3}{A^+_3+A^-_3} 
\end{eqnarray}
Based on this new flux, new relative velocities are defined on the faces of the cell: 
\begin{eqnarray}
\hat{u}_i & = & \frac{\Delta x_i \tilde{F}_i}{\rm V}  \;\;
\end{eqnarray}
where $\Delta x_i$ is the grid spacing and ${\rm V}= \Delta x_1 \Delta x_2 \Delta x_3$ is the volume of the cell.  Away from the ship hull, $\hat{u}_i$ is the relative velocity plus a corrective term to conserve mass.    Inside the ship hull, $\hat{u}_i=0$ because $A_i=0$.   Near the ship hull, $\hat{u}_i$ is scaled by the fraction of area that is cut by the presence of the ship hull.   $\hat{u}_i$ is continuous across the faces of the cells along $x-$ and $y-$axes, but discontinuous across the faces along the $z-$axis because in this particular example that is the axis where the flux has been corrected.

Equation \ref{vof} is operator split.   A dilation term is added to ensure that the volume fraction remains between $0 \leq \phi \leq 1$ during each stage of the splitting \cite{puckett97}.  The resulting discrete set of equations for the first stage of the time-stepping procedure is provided below:    
\begin{eqnarray}
\tilde \phi      =  \phi^k    
& - & \frac{{\cal F}_1\left[ \left(\hat{u}^+_1\right)^k,\phi^k,\Delta t \right] 
       -          {\cal F}_1\left[ \left(\hat{u}^-_1\right)^k,\phi^k,\Delta t \right]}{\rm V} \nonumber \\
& + &  \Delta t  \tilde{\phi} \frac{\left(\hat{u}^+_1\right)^k-\left(\hat{u}^-_1\right)^k}{\Delta x_1}   \nonumber \\
\tilde{\tilde{\phi}}  = \tilde \phi   
& - & \frac{{\cal F}_2 \left[ \left(\hat{u}^+_2\right)^k,\tilde \phi,\Delta t \right]
       -          {\cal F}_2 \left[ \left(\hat{u}^-_2\right)^k,\tilde \phi,\Delta t \right]}{\rm V} \nonumber \\
& + & \Delta t \tilde{\tilde{\phi}} \frac{\left(\hat{u}^+_2\right)^k-\left(\hat{u}^-_2\right)^k}{\Delta x_2}   \nonumber \\
\phi^* =  \tilde{\tilde{\phi}} 
& - & \frac{{\cal F}_3 \left[ \left(\hat{u}^+_3 \right)^k,\hat \phi,\Delta t \right] 
      -           {\cal F}_3 \left[ \left(\hat{u}^-_3 \right)^k,\hat \phi,\Delta t \right]}{\rm V} \nonumber \\
& + & \Delta t \tilde{\tilde{\phi}} \frac{\left(\hat{u}^+_3\right)^k-\left(\hat{u}^-_3\right)^k}{\Delta x_3} \;\; ,
\end{eqnarray}
${\cal F}_i$ denotes VOF advection based on the uncut areas of the faces of the cell.   As an example, for a cell that is full of water,  ${\cal F}_1(\hat{u}_1,\phi,\Delta t )=\phi \hat{u}_1 \Delta t \Delta x_2 \Delta x_3$.   The dilation term is treated explicitly in the first two parts of the operator-slip algorithm and implicitly in the last part of the preceding equation.  Note that the order of the splitting is alternated from time step to time step to preserve second-order accuracy.

\subsection{Interface Visualization}

The free-surface interface that is reconstructed from the volume fractions is most often calculated and visualized using commercial codes.     Specifically,  commercial codes calculate the 0.5 isosurface of the volume-fraction function $\phi$.   The free-surface interface that is calculated from the 0.5  isosurface is different from the free-surface interface that is reconstructed from the volume fractions.   To illustrate this point, consider a cell whose volume fraction $\phi_o$ is between half full and full, $0.5 \leq \phi_o \leq 1$.   Let $\Delta z$ denote the height of the cell.    Assume that the free-surface interface is horizontal and that all the fluid is sitting in the bottom of the cell.    Then the height of the free-surface interface above the bottom of the cell based on VOF reconstruction is as follows: 
\begin{eqnarray}
\label{eqn:vof1}
\eta = \phi_o \Delta z \;\; .
\end{eqnarray}
\noindent In contrast, if the cell above is filled with air. then based on the 0.5 isosurface, the height of the free-surface interface is
\begin{eqnarray}
\label{eqn:vof2}
\eta = \left( \frac{3}{2} - \frac{1}{2 \phi_o} \right) \Delta z \;\; .
\end{eqnarray}
\noindent The maximum difference between equations \ref{eqn:vof1} and \ref{eqn:vof2} occurs when $\phi_o=3/4$.  The error at this point is about 11\% higher for 0.5 isosurface relative to VOF reconstruction.  If the volume fraction is less than $\phi < 0.5$, then the 0.5 isosurface does not even exist.   This is problematic in visualizations of turbulent flows with lots of spray because droplets and sheets of water can suddenly appear and disappear.

\subsection{Radiation Conditions}

Exit boundary conditions are required in order to conserve mass and flux.   For ships with forward speed, an Orlanski-like formulation \cite{orlanski93}  provides the necessary radiation condition.   
\begin{eqnarray}
\label{eqn:orlanski}
\frac{\partial u_1}{\partial t} + u_c \frac{\partial u_1}{\partial x}  =  0 \;\; .
\end{eqnarray}
\noindent $u_c$ is the forward speed of the ship, and $u_1$ is the water-particle velocity along the x-axis.    For the other components of velocity and the volume fraction, zero gradients are imposed at the exit of the computational domain:
\begin{eqnarray}
\frac{\partial u_2}{\partial x}=\frac{\partial u_3}{\partial x}=\frac{\partial \phi}{\partial x} = 0 \;\; .
\end{eqnarray}
\noindent Neumman conditions are specified for the pressure in a manner that is very similar to the imposition of free-slip conditions on the ship hull  (see equations \ref{eqn:integral1} thru \ref{eqn:integral2}).   Based on the x-component of momentum, 
\begin{eqnarray}
\frac{1}{\rho}\frac{\partial P}{\partial x} =-\frac{\partial u_1}{\partial t} +R_1  \;\; .
\end{eqnarray}
Upon substitution of equation \ref{eqn:orlanski} into the preceding equation, the following Neumann condition is derived for the pressure at the exit of the computational domain:
\begin{eqnarray}
\label{eqn:exit}
\frac{1}{\rho} \frac{\partial P}{\partial x} =  u_c \frac{\partial u_1}{\partial x} + R_1 \;\; .
\end{eqnarray}
This equation is substituted into the set of finite-volume equations that govern the pressure (see equation \ref{eqn:integral2}).  

Equations  \ref{eqn:orlanski} thru \ref{eqn:exit} prevent the reflection of disturbances back into the interior of the computational domain.   However, these equations do not guarantee the conservation of mass.   In order to conserve mass, a regridding procedure is introduced.   The initial volume fraction is integrated for the grid cells that are on the leading edge of the computational domain.    This integrated quantity is used to maintain a constant mean water level at the entrance to the computational domain.    At the end of each time step, changes in the integrated volume fraction are calculated.     Any changes in the integrated volume fraction are eliminated by imposing a vertical velocity that brings the mean water level at the leading edge back into alignment.   The velocity correction is used to move the  volume fractions over the entire computational domain either up or down, depending on the situation.  A VOF method is used to move the volume fractions.   The VOF method ensures that the free-surface interface remains sharp during the regridding process.

\subsection{Initial Transients}

Initial transients are minimized using an adjustment procedure.   An analysis of adjustment procedures as it applies to free-surface problems is provided in \citeasnoun{dommermuth94} and \citeasnoun{dommermuth00}.  Let $f(t)$ denote the adjustment factor as a function of time, then $f(t)$ and its derivative $f'(t)$ are by definition
\begin{eqnarray}
f(t) & = & 1-\exp(-(\frac{t}{T_o})^2) \nonumber \\ 
f'(t) & = & 2 \frac{t}{T^2_o} \exp(-(\frac{t}{T_o})^2)  \;\; ,
\end{eqnarray}
where $T_o$ is the adjustment time.   The adjusted velocity of a ship moving with unit forward speed along the x-axis is
\begin{eqnarray}
v_1 = f(t) \;\; .
\end{eqnarray}
For a ship hull that is oscillating up and down, the vertical motion ($z$) and vertical velocity ($v_3$)  of the free surface in a body-fixed coordinate system are 
\begin{eqnarray}
\label{eqn:heave}
z & = & A sin(\omega t) f(t) \nonumber \\
v_3 & = & A \omega \cos(\omega t) f(t) + A sin(\omega t) f'(t) \;\; .
\end{eqnarray}
$A$ is the amplitude and $\omega$ is the frequency of the vertical motion. Since the free surface moves relative to the ship hull, there is no need to recalculate how the ship hull intersects the cartesian grid.  This is the major advantage of body-fixed coordinate systems relative to coordinate systems that are not fixed relative to the body \cite{repetto}.  The main disadvantage of body-fixed coordinate systems is that for rotational modes of motion, the Courant condition may be very restrictive near the edges of the computational domain.

\subsection{Enforcement of Courant Conditions}

The momentum equations are integrated in time using an explicit Runge-Kutta algorithm.    As a result, a Courant condition must be enforced for the maximum relative velocity:
\begin{eqnarray}
\left| u_i -v_i \right| \leq C \frac{\Delta x_i}{\Delta t}
\end{eqnarray}
$C$ is a coefficient that ensures that the Courant condition is satisfied for both the momentum equations and the VOF advection.   Typically, $C=0.45$ in the numerical results that are presented in this paper.   If the Courant condition is exceeded, the magnitude of the velocity is reduced such that the Courant condition is satisfied.  This clipping of the velocity field tends to occur in regions where fine spray is formed, especially in the rooster-tail region.

\subsection{Treatment of convective terms}

The convective terms  in the momentum equations (see Equation \ref{navi}) are calculated using a slope-limited, QUICK, finite-difference scheme \cite{leonard97}.  Special treatments are required near the ship hull.   One possibility is to use one-sided differencing.   However, one-sided differencing is often unstable.    Another possibility is to extend the velocity of the fluid into the ship hull.    In this case, setting the velocity equal to zero inside the body is stable, but too ``sticky."   Another possibility is to extend the fluid velocity into the ship hull in such a manner that the no-flux condition is met right at the ship hull.     The interior flow that meets this condition is as follows:
\begin{eqnarray}
u_i = \left( v_j n_j \right) n_i \;\; ,
\end{eqnarray}
where recall that $v_j$ is velocity of the body and $n_j$ is the unit normal that points along gradient of the signed-distance function ($\psi$).   At the ship hull, $u_i n_i = v_i n_i$ using this formulation of the interior flow.

\subsection{Density Smoothing}

The density as a function of the volume fraction is smoothed using a three-point stencil (1/4,1/2,1/4) that is applied consecutively along each of the cartesian axes.   This improves the conditioning of the Poisson equation (Equations \ref{eqn:poisson1} \& \ref{eqn:poisson2}).   If the density is smoothed, then the same smoothed density must be used in the projection steps (Equations \ref{eqn:runge1} \& \ref{eqn:runge2}).

\section{Results} \label{sec:results}

\subsubsection{5365 geometry (Athena)}

Experimental measurements of model 5365 have been performed at Froude numbers $Fr=0.2518$ and $0.4316$ that correspond to equivalent full-scale speeds of 10.5 and 18 knots, respectively.    Details of the experimental measurements are provided in \citeasnoun{wilson06}.

Corresponding to these experiments, three-dimensional numerical simulations using 680x192x128=16,711,680 grid points, 4x8x4=128 sub-domains, and 128 nodes have been performed on a Cray XT3.  The length, width, depth, and height of the computational domain are respectively 3.0, 1.0, 1.0, 0.5 ship lengths (L).  Grid stretching is employed in all directions.  The smallest grid spacing is 0.002L near the ship and mean waterline, and the largest grid spacing is 0.02L in the far field.  This provides about 8 x 61 cells across the transom of the low Froude-number case, and 11 x 61 grid cells for the high Froude-number case.   For the low Froude-number case, there are 200 cells per transverse wavelength (0.398L) where the grid spacing is fine and 20 grid cells where it is coarse.   For the high Froude-number case, there are 585 cells per transverse wavelength (1.17L) where the grid spacing is fine and 58.5 grid cells where it is coarse.  Initial transients are minimized by slowly ramping up the free-stream current.  The period of adjustment associated with this ramp up is 0.5 in non-dimensional units of time, where $T=(L/g)^{1/2}$ is the normalization factor.   For these simulations, the non-dimensional time step is t=0.0005.  The numerical simulations run 12001 time steps corresponding to 6 ship lengths.  They each require 50 hours of wall-clock time.

\begin{figure*}
\begin{center}
\includegraphics[width=\linewidth]{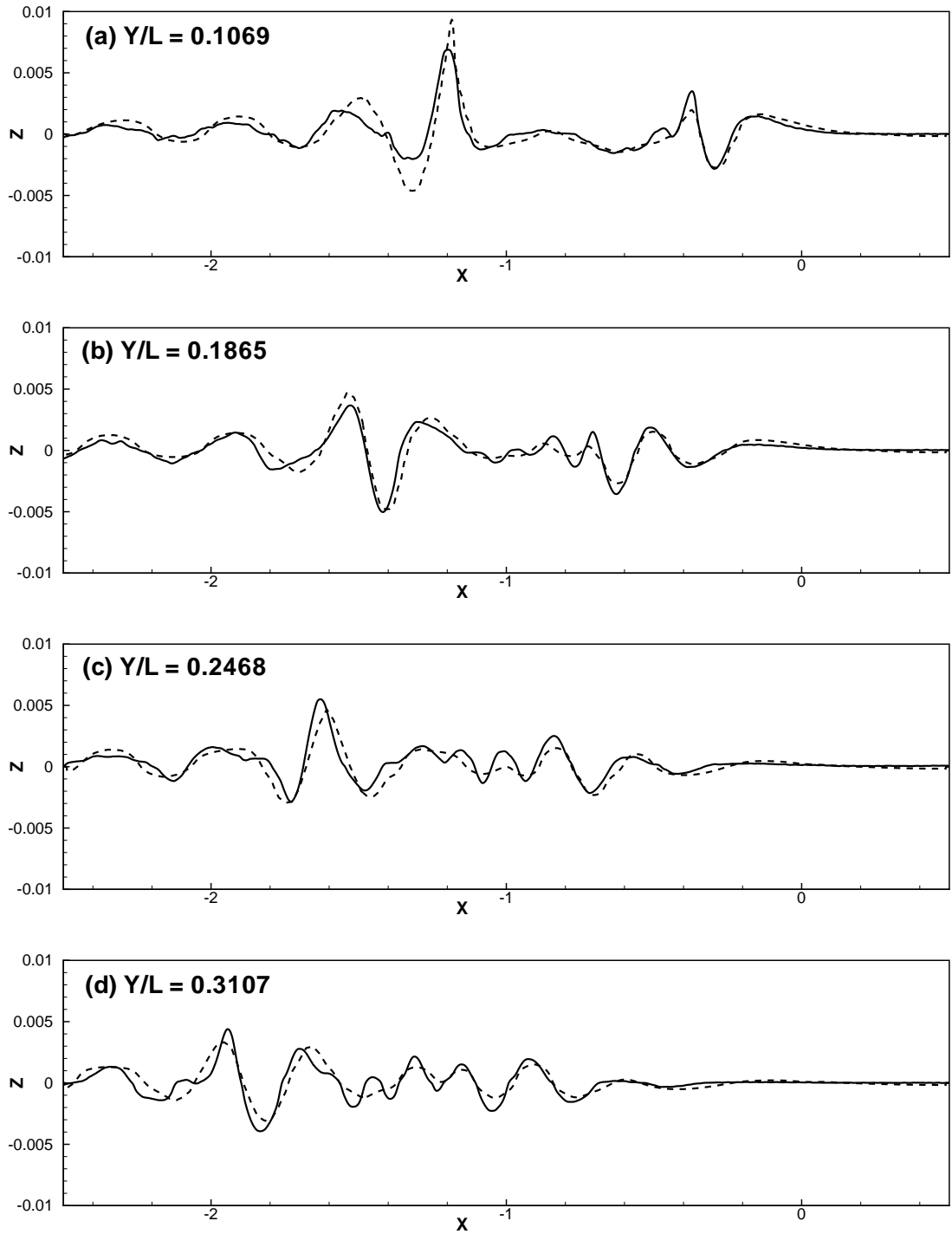}
\caption{\label{athena_105} Model 5365 (Athena) 10.5 knot wave cuts.}
\end{center}
\end{figure*}

Figure \ref{athena_105} shows wave cuts for the 10.5 knot case.  The correlation coefficients between experimental measurements and numerical predictions for parts (a) thru (d) of Figure \ref{athena_105}  are 0.89, 0.91, 0.85, and 0.86, respectively.  The solid and dashed lines respectively denote the experimental measurements and the numerical predictions.  The correlation gets poorer in the region where the grid spacing along the y-axis gets poorer.   The shortest waves are not resolved by the numerical simulations.  More grid resolution is required.   Convergence studies are in progress.

\begin{figure*}
\begin{center}
\includegraphics[width=\linewidth]{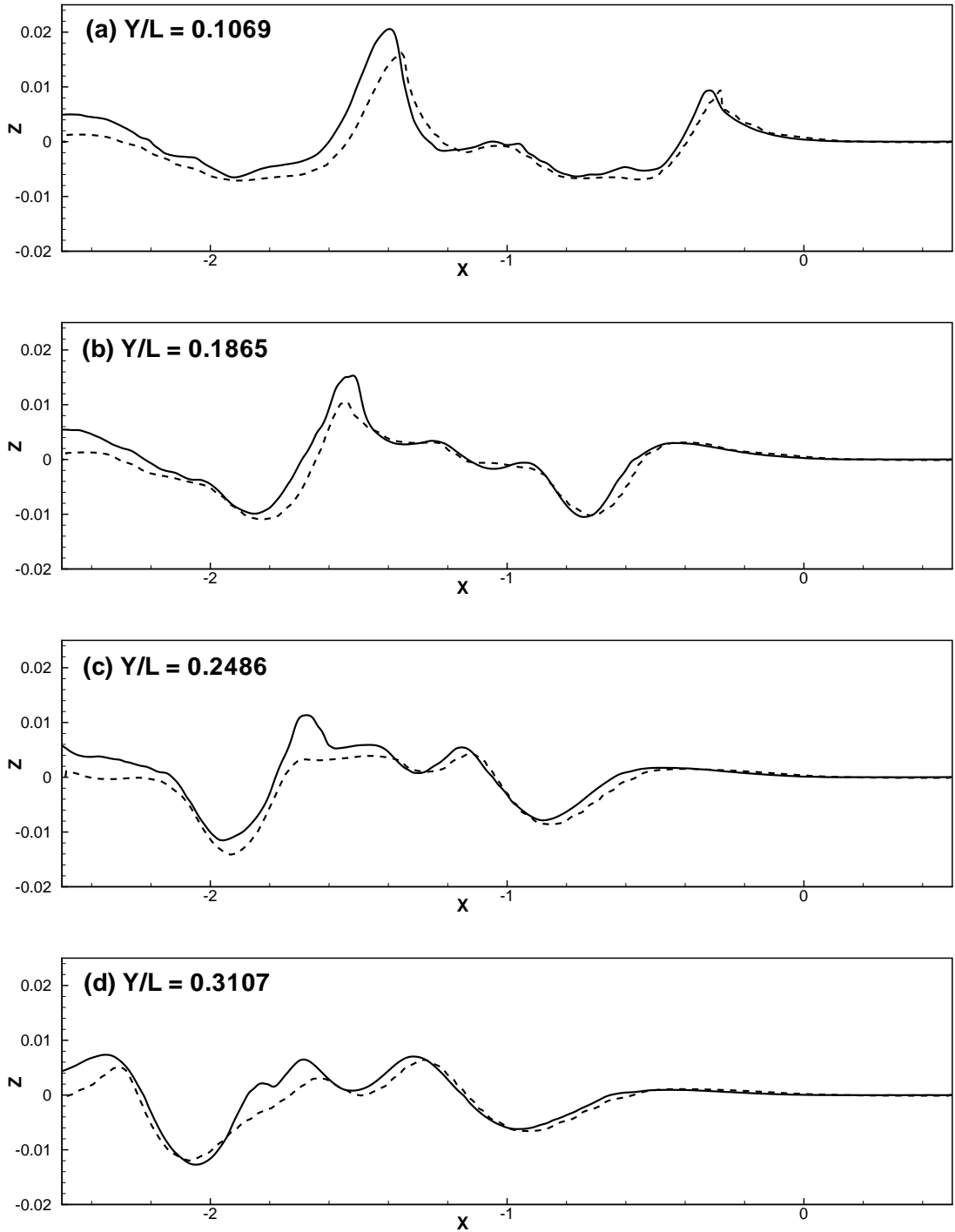}
\caption{\label{athena_18} Model 5365 (Athena) 18 knot wave cuts.}
\end{center}
\end{figure*}

Figure \ref{athena_18} shows wave cuts for the 18 knot case.  The correlation coefficients between experimental measurements and numerical predictions for parts (a) thru (d) of Figure \ref{athena_18}  are 0.89, 0.92, 0.88, and 0.91, respectively.   In general, the high Froude-number simulation is in slightly better agreement with the experimental measurements than the low Froude-number simulation, probably because the waves are longer.    However, both simulations would benefit from using higher resolution, especially near the bow  and transom where there is wave breaking and flow separation.

\subsubsection{5415 geometry}

Experimental measurements of a DDG model 5415 have been performed at Froude numbers $Fr$=0.2755 and $0.4136$ that correspond to equivalent full-scale speeds of 20 and 30 knots, respectively.   The measurements include free-surface profiles on the ship hull, free-surface elevations near the bow and stern using a whisker probe, and total drag.  The length, beam, and draft of the model are respectively 5.72m, 0.388m, and 0.248m.   The model-scale speeds are 4.01 and 6.02 knots.  Details of the hull geometry, including the sinkage and trim, are provided by the \citeasnoun{ddg5415}.

A three-dimensional numerical simulation using 800x192x192=29,491,200 grid points, 4x8x8=256 sub-domains, and 256 nodes has been performed on a Cray XT3.  The length, width, depth, and height of the computational domain are respectively 3.0, 1.0, 1.0, 0.5 ship lengths (L).  Grid stretching is employed in all directions.  The smallest grid spacing is 0.0008L near the ship and mean waterline, and the largest grid spacing is 0.05L in the far field.  For the high Froude-number case, this provides about 15 x 100 grid cells across the transom and 1340 grid cells per transverse wavelength (1.07L) where the grid spacing is fine and 20 grid cells where it is coarse.  Initial transients are minimized by slowly ramping up the free-stream current.  As before, the period of adjustment associated with this ramp up is 0.5 in non-dimensional units of time. For this simulation, the non-dimensional time step is t=0.0002.  The numerical simulation runs 28001 time steps corresponding to 5.6 ship lengths.  It requires 125 hours of wall-clock time.

\begin{figure*}
\begin{center}
\includegraphics[width=0.60\linewidth]{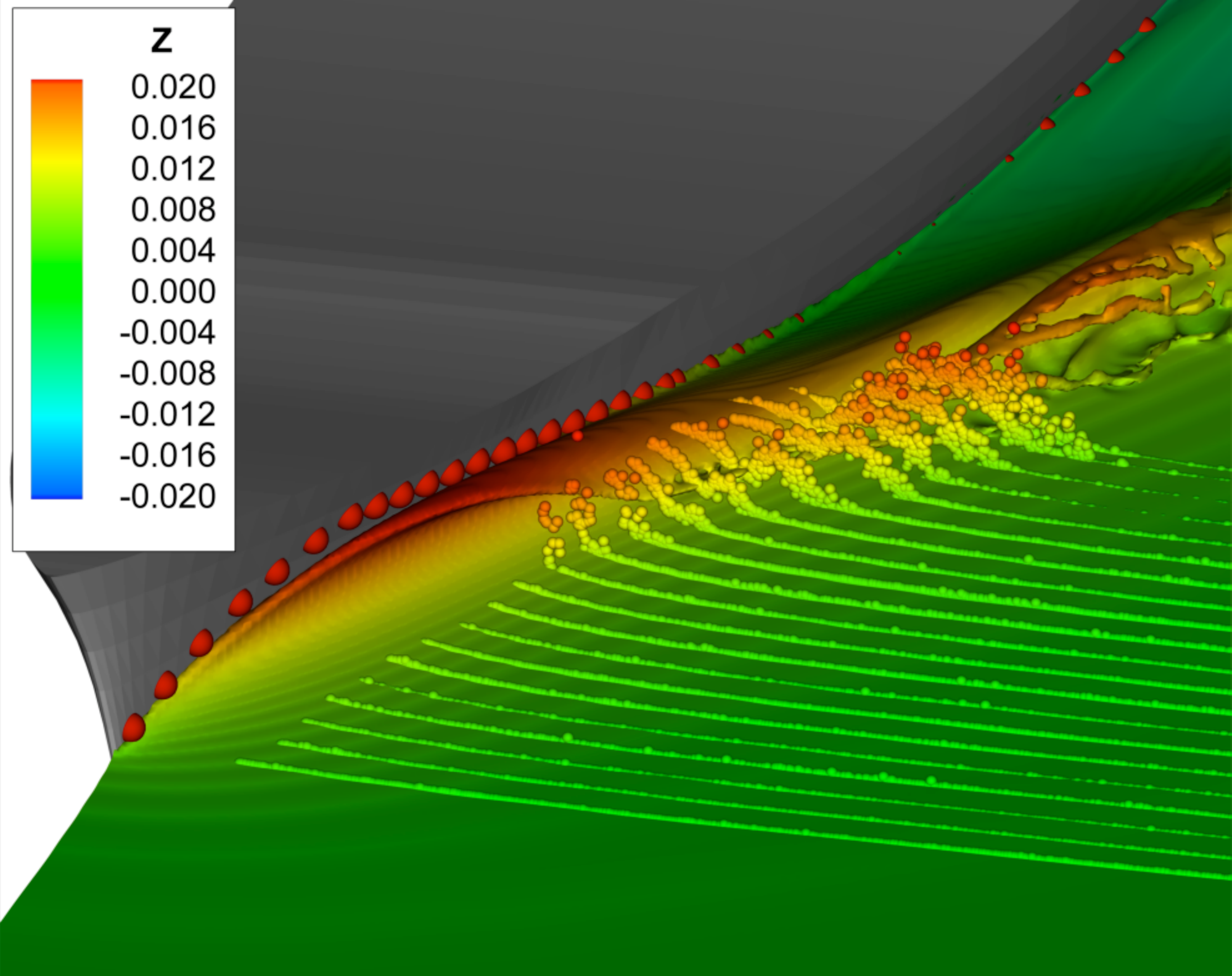}
\caption{\label{ddg_bow} Model 5415 bow view.}
\end{center}
\end{figure*}

Figure \ref{ddg_bow} compares NFA predictions to experimental measurements for the flow near the bow.   The free-surface profile measurements are denoted by spherical symbols along the ship hull.   The ship hull is outlined in grey.   Whisker-probe measurements are indicated by the small spherical symbols transverse to the ship.  Due to the measuring technique, whisker-probe data provides an upper bound of the free-surface elevation.   NFA predictions of the free surface are denoted by the color contour.  In general, the whisker-probe measurements agree well with the upper bound of the free-surface predictions.  NFA correctly predicts the overturning of the bow wave and the resulting splash up slightly aft of the bow.  At this resolution, the numerical simulations do not  resolve the very thin sheets which characterize the run-up near the bow.  As a result, the NFA predictions are slightly lower than the maximum free-surface profile that has been measured.

\begin{figure*}
\begin{center}
\includegraphics[width=0.60\linewidth]{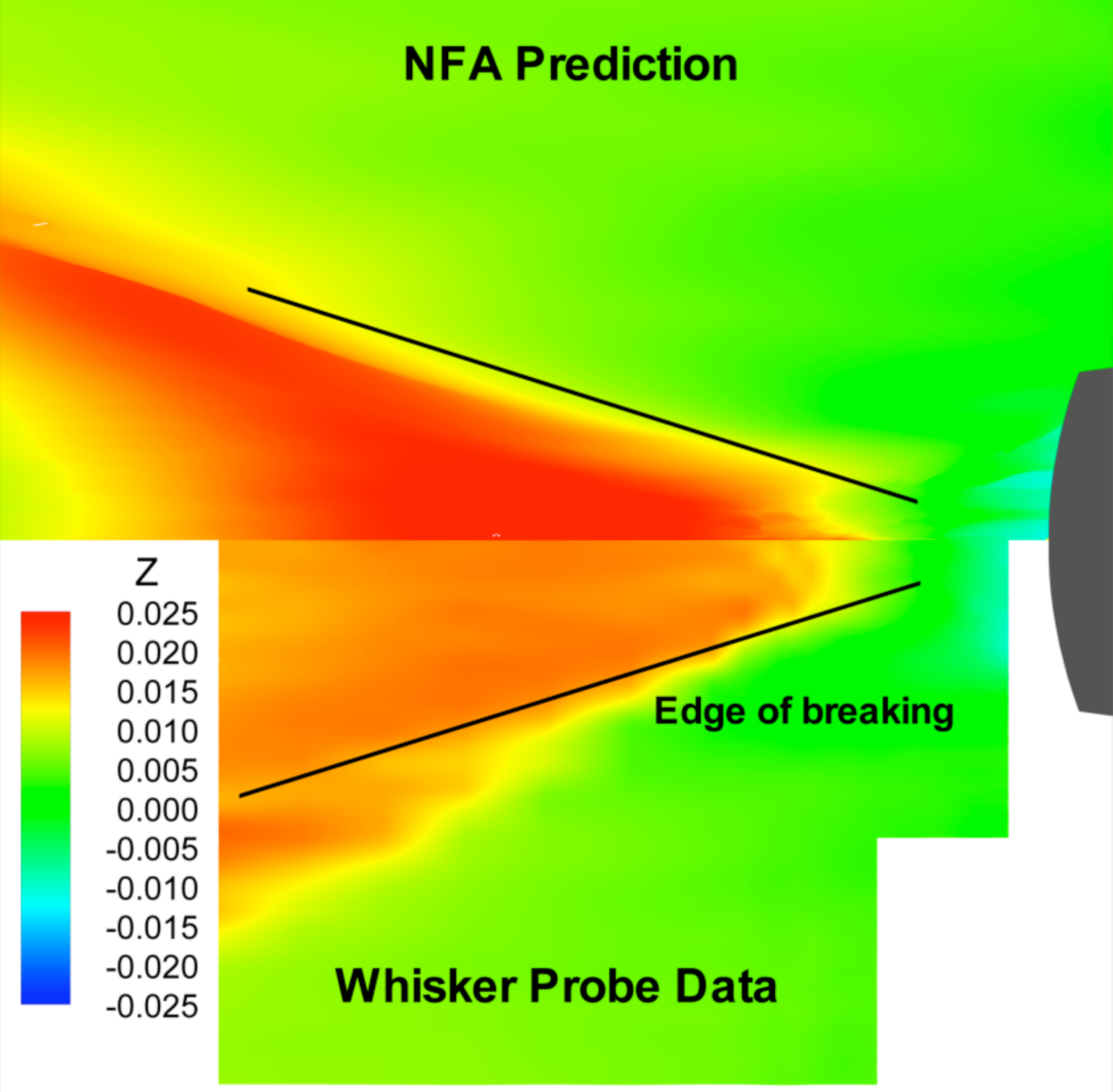}
\caption{\label{ddg_stern} Model 5415 stern view.}
\end{center}
\end{figure*}

Figure \ref{ddg_stern} compares NFA predictions to whisker-probe measurements for the flow near the stern.  The portion above the centerline of the ship represents NFA results while the portion below is based on experiments.  Black lines mark the edges where spilling occurs. NFA accurately captures the flow separation from the transom stern and agreement between predictions and measurements is good overall.  However, at this resolution some spilling along the edges of the rooster tail is not captured. As a result, the predicted rooster-tail amplitude directly astern of the transom is higher than measurements.  Simulations that resolve the breaking in this region may provide the dissipation of energy that is necessary to reduce the wave amplitude in the rooster-tail region.

\begin{figure*}
\begin{center}
\includegraphics[width=\linewidth]{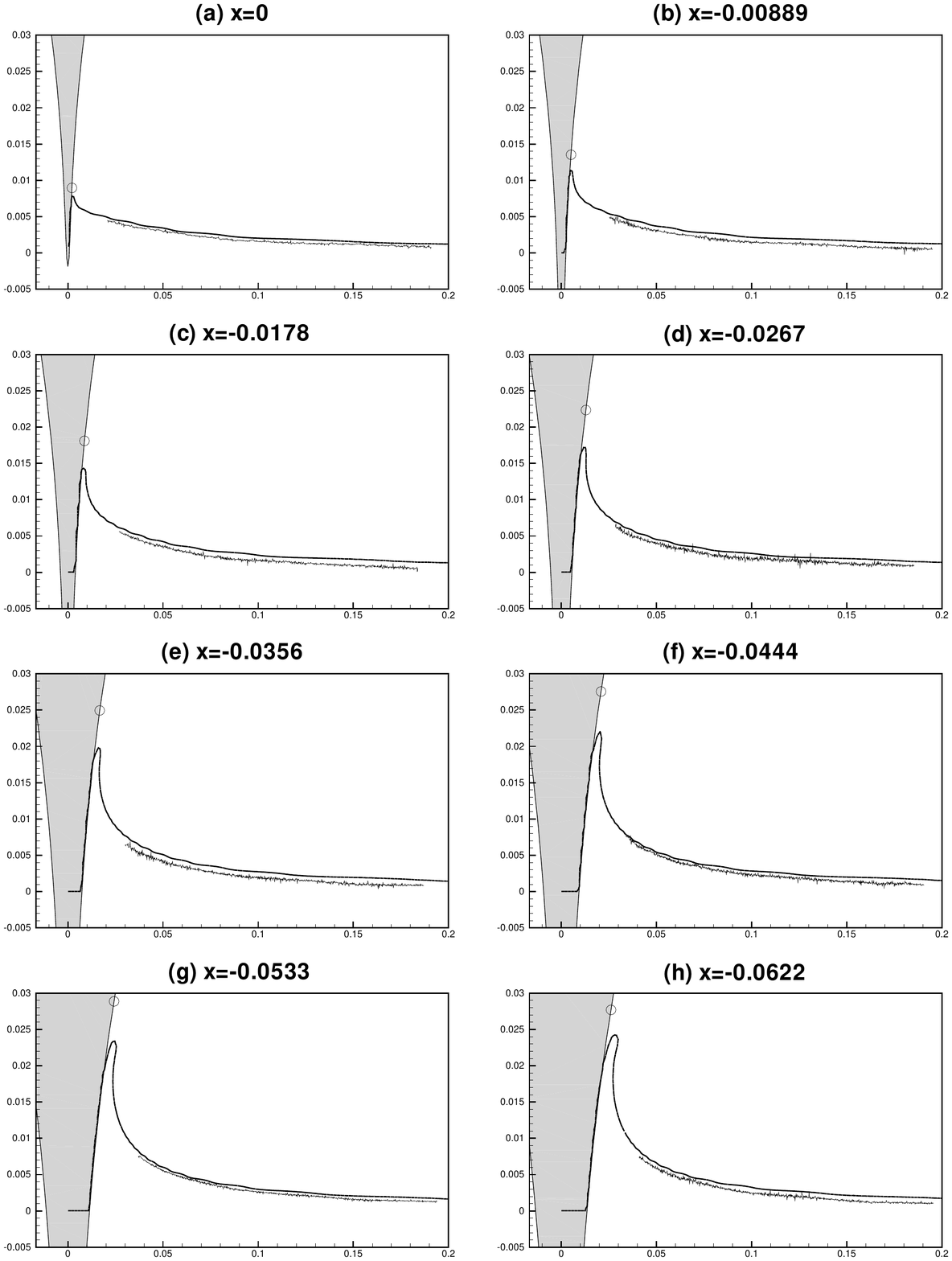}
\caption{\label{ddg_whisker} Model 5415 transverse wave cuts.}
\end{center}
\end{figure*}

Figure \ref{ddg_whisker} shows transverse cuts of the free-surface elevation near the bow. The cross section of the port side of the hull is outlined using a grey shade.  Circular symbols denote the profile measurements along the side of the hull.  Solid lines denote whisker-probe measurements.   Dashed lines denote NFA predictions.  Results are shown for various stations aft of the bow from (a) x=0 to (t) x=-0.169L, where L is the ship length. The figures show the overturning of the bow wave.  The initial onset of air entrainment is evident in the NFA predictions.  In addition, fragments of splash up are also captured.    As expected, the whisker-probe measurements provide an upper envelope to the numerical predictions.  This effect is illustrated in Figure \ref{ddg_whisker}k, where the bow wave is just beginning to overturn.   For this station, there is a sudden jump in the whisker-probe measurement that corresponds to the instrument measuring the top of the spray sheet.  The envelop of the plunging event is captured well by the numerical simulations as illustrated in Figures \ref{ddg_whisker}k-o.   Figures \ref{ddg_whisker}o \& \ref{ddg_whisker}p show the initial stages of splash up.   The results of the numerical simulation are not as energetic as the experiments.    Higher resolution may help in this particular region.   The entrainment of air is observed in Figure \ref{ddg_whisker}p-t.   The splash up is resolved better by the numerical simulations in Figures  \ref{ddg_whisker}s \& \ref{ddg_whisker}t.   However, the whisker-probe measurements are consistently lower than the numerical simulations in Figures \ref{ddg_whisker}q-t.   This requires further study to find the source of the discrepancy.

Figure \ref{ddg_drag} shows the total resistance as a function of time.  Time is normalized by $L_o/U_o$, where $L_o=5.72m$ and $U_o=3.10 m/s$.   The solid line is the measured steady-state value and the dashed line is the unsteady NFA prediction.  The NFA predictions are calculated by integrating the normal component of the pressure along the direction of travel over the surface of the ship hull.  Based on ITTC line, the portion of the drag that tangential to the ship is added to these results to predict the total drag.  This additional term is required because NFA is an Euler code that does not directly predict skin-friction drag.  Unlike potential-flow formulations, NFA does not require a correction for residuary resistance associated with the shedding of vorticity because Euler formulations account for base drag.  The figure shows NFA results converging to the steady-state resistance.  We note that there are known long-time transients associated with a ship accelerating from rest to constant forward speed \cite{dommermuth04}.  

\subsubsection{Sphere geometry}

Consider the motion of a heaving sphere that is moving with forward speed.   The purpose of this study is to build toward developing a capability that is suitable for forced-motion studies and seakeeping.   The Froude number is $Fr=U_o/\sqrt{g D}=0.5$, where $D$ is the diameter of the sphere.   The normalized diameter of the sphere is $D=1$.  The amplitude of the heaving motion is $A=0.25$ and the frequency of oscillation is $\omega=2 \pi$ (see Equation \ref{eqn:heave}).  Three grid resolutions are studied: $64^3=262,144$, $128^3=2,097,152$, and $256^3=16,777,216$ grid points.  The length, width, and height of the computational domain are 4.  The smallest grid spacings for the coarse, medium, and fine grids are respectively 0.00412, 0.0206, and 0.0103.  The time steps are respectively 0.0025, 0.00125, and 0.0006225.

Figure \ref{sphere_forx} shows the x-component of force acting on a sphere moving with forward speed, and Figure \ref{sphere_forz} shows the z-component.   The forces are normalized by the displacement of the sphere, which is initially half immersed.   The forces only include the portion of the pressure that is directed normal to the surface of the sphere.   The effects of skin friction are not included.   First- and second-harmonic interactions are evident in both components of force.  The second-harmonic interactions are particularly strong for the vertical component of force, probably due to wave breaking and collapse beneath the sphere.  

Figure \ref{sphere_sequence} shows a time sequence of a heaving sphere moving with forward speed.   The initial start-up stage is shown.  A plunging breaker forms near the bow in Figure \ref{sphere_sequence}g.   Upon breakup, the flow becomes very turbulent.

\section{Conclusions}

In terms of progress, it is interesting to consider the results of research reported in earlier ONR symposiums.   \citeasnoun{dommermuth98} study the flow near the bow of model 5415 using a variable-density, cartesian-grid formulation.   A body force is used by \citeasnoun{dommermuth98} to impose the body boundary condition.   The numerical results of     \citeasnoun{dommermuth98}  barely capture the initial onset of wave overturning near the bow.  \citeasnoun{sussman01} continue to develop interface capturing methods.  Once again, comparisons are shown to the bow flow of model 5415.   The results do not show significant improvement over their earlier results.  However, their calculations of the breakup of a turbulent spray sheet illustrate a novel application of interface-capturing methods.  \citeasnoun{dommermuth04} use two methods to study the flow around model 5415, a vertical strut, and a bluff wedge.  The first method uses free-slip conditions on the hull in combination with a hybrid level-set and VOF interface-capturing method.   In addition,  adaptive mesh refinement (AMR) is used to improve grid resolution near the hull and free-surface interface.   Their preliminary results illustrate the efficiency  of AMR.  The second method uses body-force and VOF formulations on a cartesian grid with no grid stretching.  The results show more fine-scale detail than the earlier studies.  The predicted free-surface elevations compare well with experiments, but the body-force method is too ``sticky" because too much fluid is dragged with the ship hull.  Based on these results, the present research uses free-slip boundary conditions to impose the body boundary condition to reduce stickiness.   The VOF algorithm has been generalized to include free-slip conditions on the ship hull.  The grid is stretched along the cartesian axes to improve grid resolution.   Together, these new formulations enable the modeling of complex free-surface flows.

It is not possible to show all of the details in the current numerical predictions through the use of figures.   In order to study the flow in even more detail, several animations have been prepared at the flow visualization center at ERDC.   The animations are accessible by contacting the authors.  Animations are available for all the cases shown in this paper.

\section{Acknowledgements}

This research is supported by ONR under contract numbers N00014-04-C-0097. Dr. Patrick Purtell is the program manager.  This work was supported in part by a grant of computer time from the DOD High Performance Computing Modernization Program (http://www.hpcmo.hpc.mil/).  The numerical simulations have been performed on the Cray XT3 at the U.S. Army Engineering Research and Development Center and the Cray X1 at the Army High Performance Computing Research Center.

\bibliography{26onr}
\bibliographystyle{26onr}

\clearpage

\addtocounter{figure}{-1}
\begin{figure*}
\begin{center}
\includegraphics[width=\linewidth]{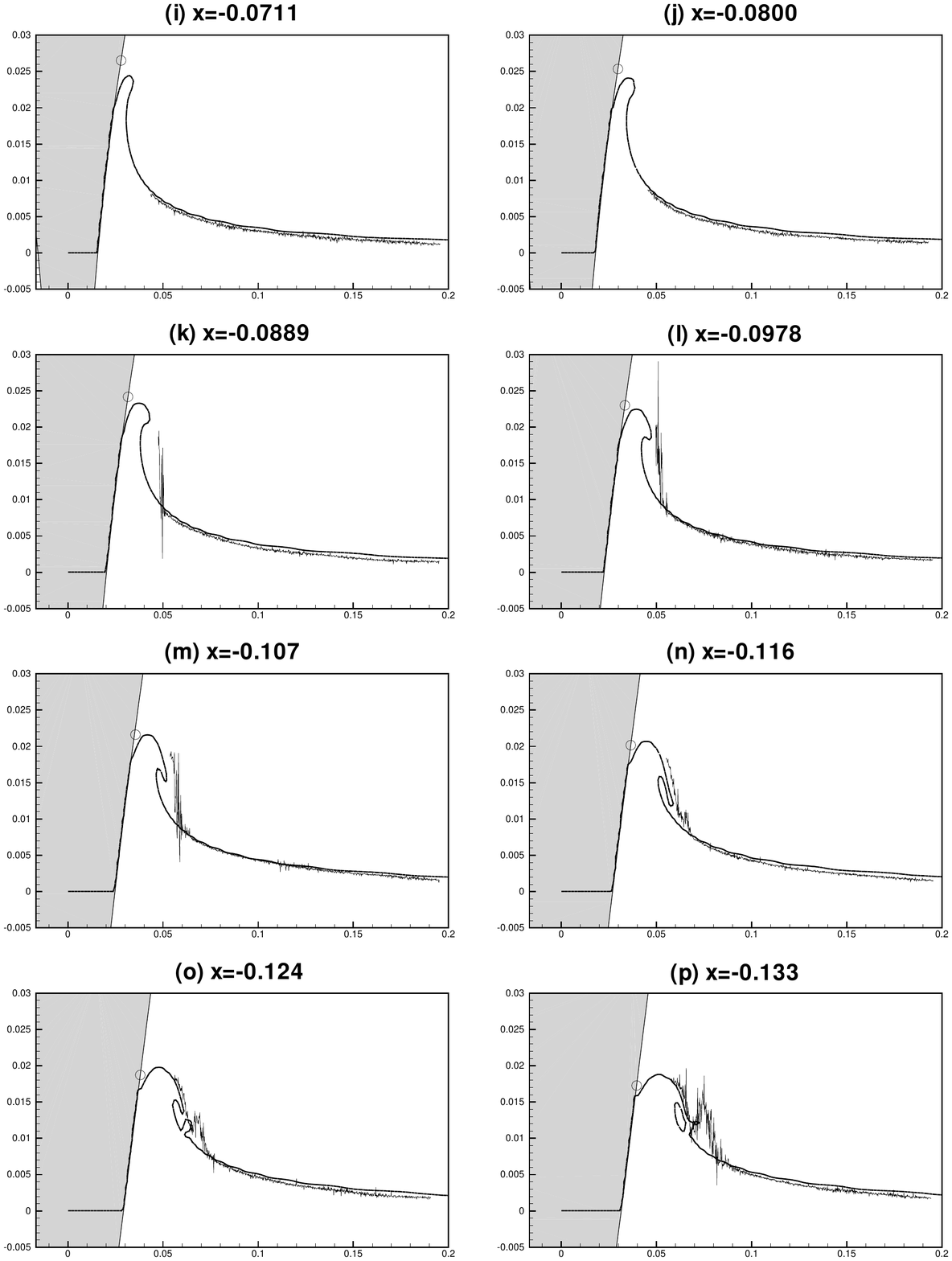}
\caption{Model 5415 transverse wave cuts, continued.}
\end{center}
\end{figure*}

\addtocounter{figure}{-1}
\begin{figure*}
\begin{center}
\includegraphics[width=\linewidth]{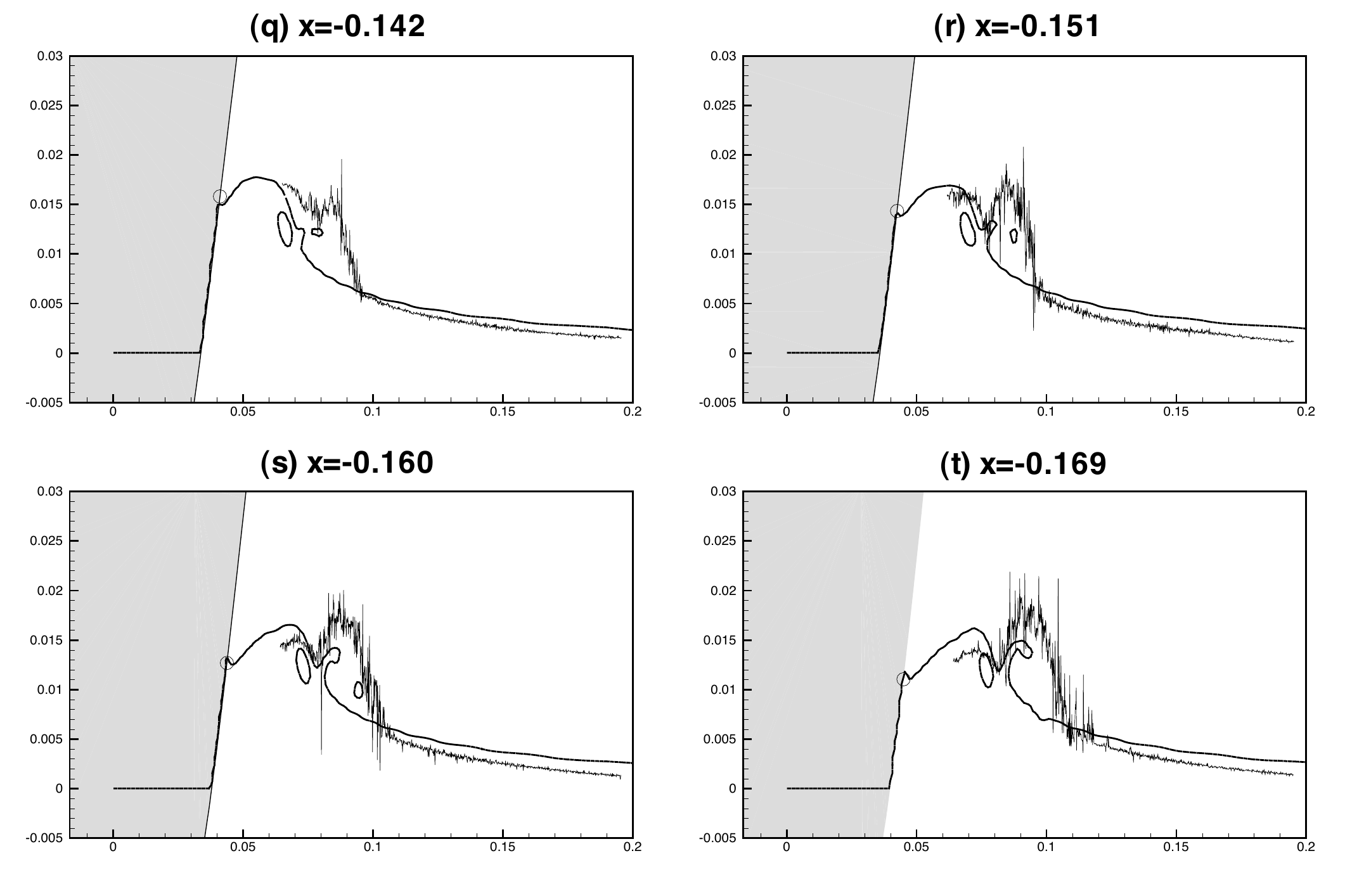}
\caption{Model 5415 transverse wave cuts, continued}
\end{center}
\end{figure*}

\begin{figure*}
\begin{center}
\includegraphics[width=0.6\linewidth]{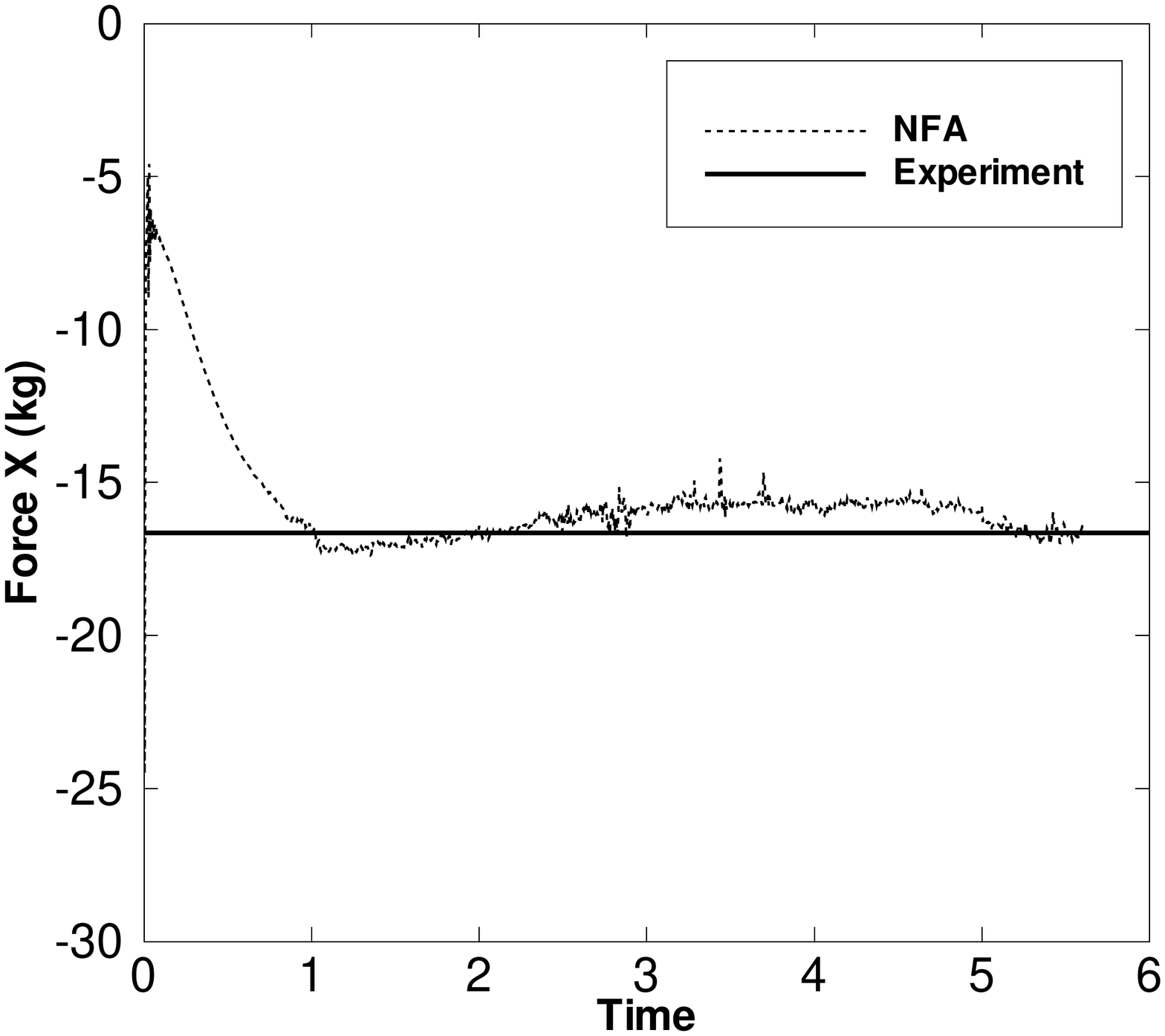} 
\caption{\label{ddg_drag} Model 5415 resistance.}
\end{center}
\end{figure*}

\begin{figure*}
\begin{center}
\includegraphics[width=0.7\linewidth]{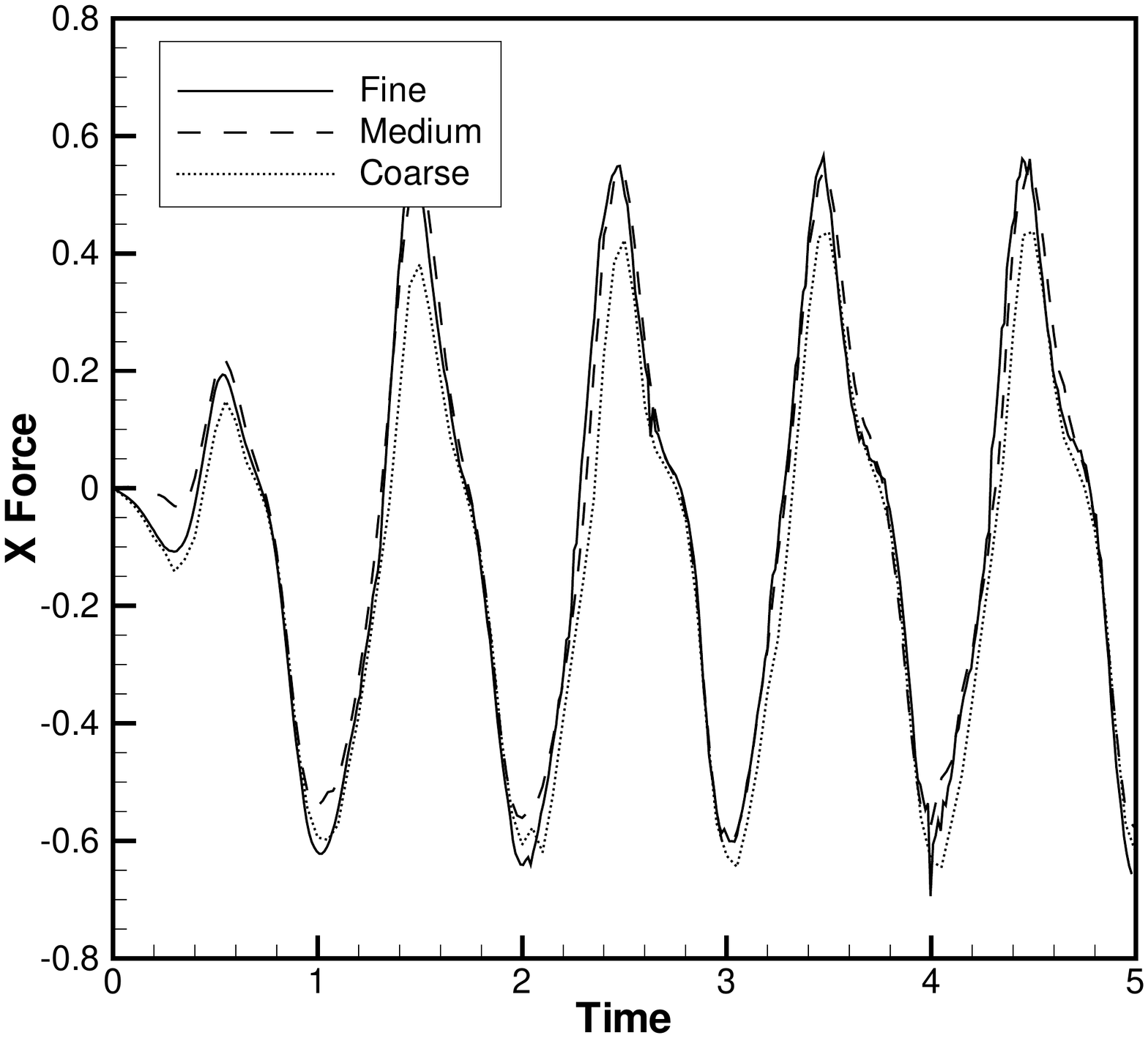} 
\caption{\label{sphere_forx} X-component of force acting on a heaving sphere moving with forward speed.}
\end{center}
 \end{figure*}

\begin{figure*}
\begin{center}
\includegraphics[width=0.7\linewidth]{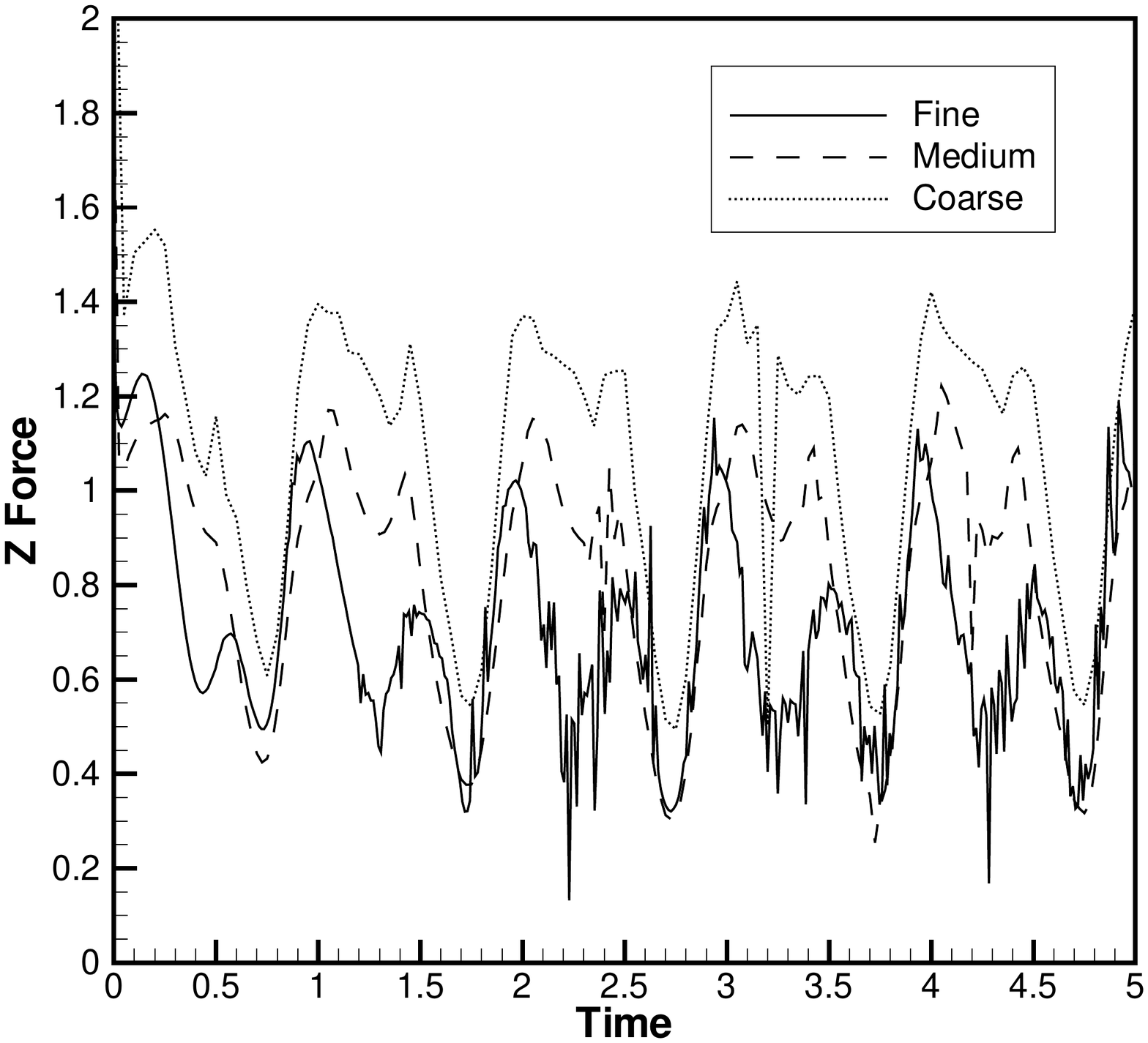} 
\caption{\label{sphere_forz} Z-component of force acting on a heaving sphere moving with forward speed.}
\end{center}
\end{figure*}

\begin{figure*}
\begin{center}
\includegraphics[width=\linewidth]{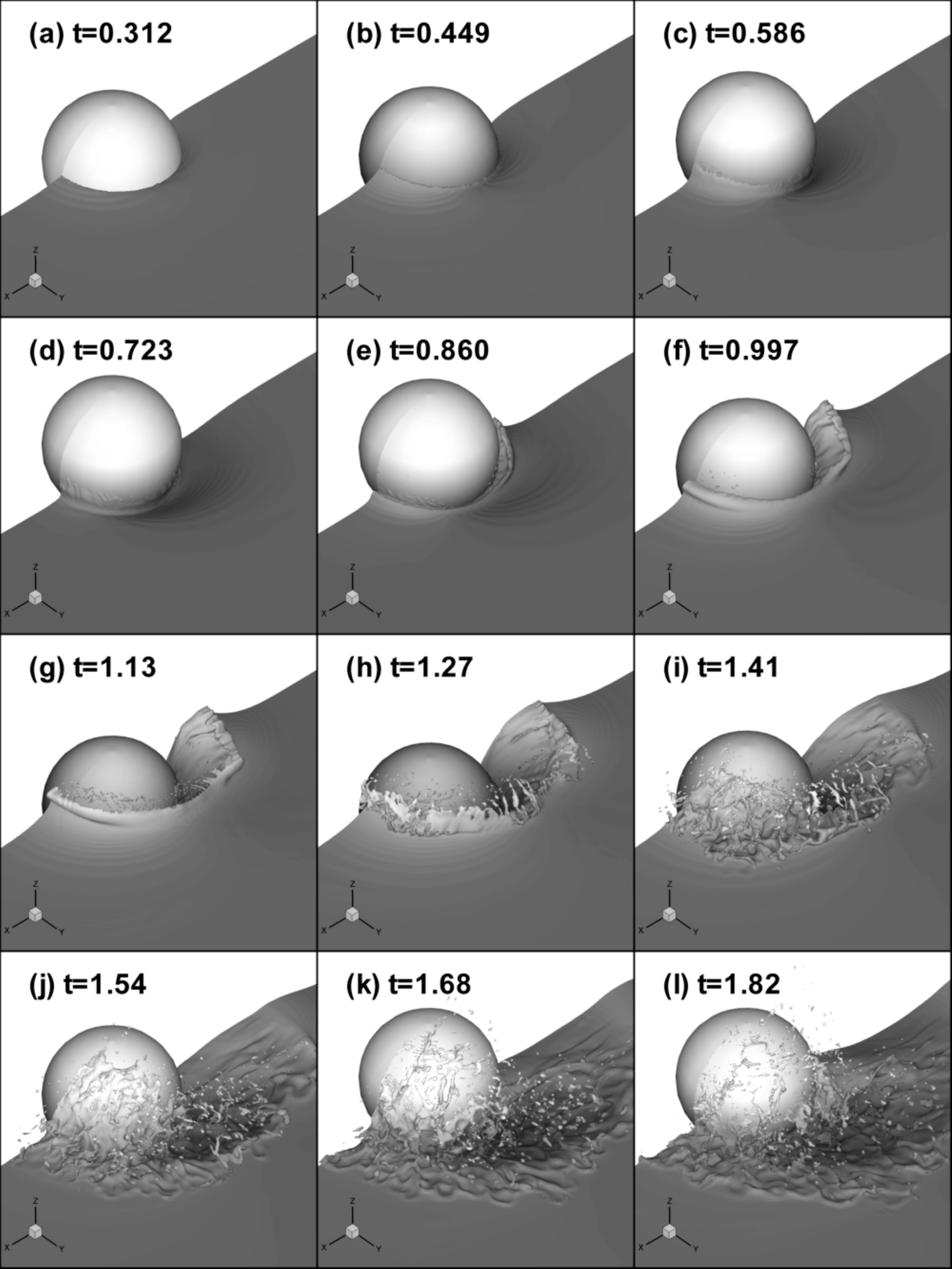} 
\caption{\label{sphere_sequence} Time sequence of a heaving sphere moving with forward speed.}
\end{center}
\end{figure*}

\end{document}